\documentclass{sigchi}

\CopyrightYear{2024}

\usepackage{balance}       
\usepackage{graphics}      
\usepackage[T1]{fontenc}   
\usepackage{txfonts}
\usepackage{mathptmx}
\usepackage[pdflang={en-US},pdftex]{hyperref}
\usepackage{color}
\usepackage{booktabs}
\usepackage{textcomp}

\usepackage{microtype}        
\usepackage{ccicons}          

\usepackage{todonotes}

\def\plaintitle{Thought Experiments in Design Fiction for Visualization}

\def\emptyauthor{}
\def\plainkeywords{Authors' choice; of terms; separated; by
  semicolons; include commas, within terms only; required.}

\makeatletter
\def\url@leostyle{%
  \@ifundefined{selectfont}{
    \def\UrlFont{\sf}
  }{
    \def\UrlFont{\small\bf\ttfamily}
  }}
\makeatother
\def\pprw{8.5in}
\def\pprh{11in}

\definecolor{linkColor}{RGB}{6,125,233}
\hypersetup{%
  pdftitle={\plaintitle},
  pdfauthor={\emptyauthor},
  pdfkeywords={\plainkeywords},
  pdfdisplaydoctitle=true, 
  bookmarksnumbered,
  pdfstartview={FitH},
  colorlinks,
  citecolor=black,
  filecolor=black,
  linkcolor=black,
  urlcolor=linkColor,
  breaklinks=true,
  hypertexnames=false
}

\usepackage{csquotes}
\usepackage{bm}
\usepackage{tikz-cd}
\tikzset{commutative diagrams/.cd, arrow style=tikz, diagrams={>=latex}}

\setcopyright{none}
\begin{document}

\title{Thought Experiments in Design Fiction for Visualization}

\numberofauthors{1}
\author{%
  \alignauthor{Swaroop Panda\\
    \affaddr{Northumbria University}\\
    \affaddr{United Kigdom}}\\
}
\maketitle









\begin{abstract}

Thought experiments are considered valuable tools in science, enabling the exploration of hypotheses and the examination of complex ideas in a conceptual, non-empirical framework. These thought experiments can be useful in design fiction for speculating future possibilities, examining existing and alternate scenarios in new ways or challenging current paradigms. In visualization, speculating future possibilities or exploring new ways of interpreting existing scenarios can provoke critical reflection and envision novel approaches. In this paper we present such thought experiments for visualization. We conceptualize and define a thought experiment to consist of a situation, a story, and a scenario. Situations are derived from different tools of thought experiments and visualization practice; a story is an AI-generated fiction based on the situation and the scenario is the grounding of the situation and story in visualization research. We present ten such thought experiments and demonstrate their utility in visualization by deriving critiques from them. 

\end{abstract}




\section{Introduction}

Design fiction, despite encountering challenges \cite{lindley2016peer}, has gained increasing popularity within the HCI community \cite{blythe2014research,markussen2013poetics,misra2023design,muller2020understanding,lindley2016pushing} due to its ability to explore speculative futures and provoke critical reflection on emerging technologies. By employing narrative-driven prototypes and scenarios, design fiction facilitates an examination of potential societal impacts, ethical considerations, and user experiences that traditional methodologies may overlook. This approach not only enriches design research and practice by envisioning innovative possibilities but also stimulates discussions about future implications and challenges. The growing popularity of design fiction in HCI reflects a broader trend towards integrating speculative and critical thinking into the design process, enhancing the discipline's ability to address complex, future-oriented issues. 

Thought experiments are an important method in scientific inquiry, though one which is less reliant on empirical observation and more inclined towards conceptual exploration and theoretical analysis \cite{sep-thought-experiment}. Rooted in the philosophical tradition, thought experiments serve as tools for explaining complex scientific concepts, probing the boundaries of existing theories, or generating novel hypotheses that sometimes are not feasible by using empirically backed methods. Unlike empirical experiments, which involve physical manipulation, data collection and observation of phenomena, thought experiments operate within the realm of imagination, allowing researchers to mentally construct hypothetical scenarios and extrapolate their implications \cite{sorensen1992thought,sep-thought-experiment}. By engaging in thought experiments, researchers can interrogate the logical coherence of their theories, identify potential flaws or inconsistencies, and stimulate innovative insights \cite{sep-thought-experiment}. Notably, thought experiments are not confined to any specific scientific discipline but find application across a broad spectrum of fields, from physics and mathematics to psychology and ethics \cite{horowitz1991thought,helm1985thought,domenech2021thought}.  An instance of an application in AI and ethics is the trolley problem, which presents the following scenario: \textit{``A train is heading towards five people, who will be killed unless a switch is thrown to divert it towards a single person instead. Is it justifiable to save five lives by deliberately causing one death?"} \cite{thomson1984trolley}

These thought experiments are well-aligned with many methods pursued in science fiction \cite{schneider2016introduction,wiltsche2021forever,seed2008companion}. A critical point of difference is as Wiltsche suggests, thought experiments target \textit{`` “physical understanding”, science fiction novels typically have “existential understanding” as their target"} \cite{wiltsche2021forever}. These thought experiments can be aligned and adapted to the methods of design fiction. Researchers have utilized these two areas interchangeably, occasionally characterizing and defining one in relation to the other. For instance, Design fiction has been thought of as ``materialized thought experiments''\cite{bleecker2022design,blythe2018research}. Blythe and Encinas report, \textit{``In Speculative Everything Dunne and Raby describe fictional worlds and thought experiments as methodological playgrounds. They note that their two favourite forms of thought experiment are the reductio ad absurdum (where a particular claim is taken to extremes in order to test it) and the counterfactual where a historical fact is changed to see what would have happened"} \cite{blythe2018research,dunne2024speculative}.

Thought experiments hold significant promise in enriching the landscape of visualization by offering a structured framework for exploring theoretical concepts \cite{chen2017pathways} and pushing the boundaries \cite{kosara2007visualization} of traditional data visualization methods.  When applied to visualization, these experiments serve as cognitive tools for conceptualizing abstract concepts and testing the limits of visual representation. By engaging in thought experiments, practitioners can probe the implications of different data interpretations \cite{hullman2011visualization} and visualization techniques, uncovering insights that may not be readily apparent through empirical observation of users alone. Moreover, thought experiments can provide a means to interrogate the underlying assumptions and biases \cite{valdez2017framework,mansoor2018data} inherent in visualization practices, fostering a deeper understanding of the foundations that underpin visualization.

In this paper, we present thought experiments for visualization. We define thought experiments as consisting of a situation, story and scenario; where the situations are derived from the different tools of thought experiments and from reflecting upon visualization practice. A story is based upon the situation and is generated by AI and a scenario based on the situation and the generated story while being grounded in visualization research. We present 10 such thought experiments in this work. Finally, we derive three generic visualization critiques, following Kosara's process and rules \cite{kosara2007visualization} from these thought experiments.

\section{Background}
\subsection{Design Fiction}

The primary methodology that we use in this paper is that of Design Fiction. The term,``Design Fiction", was first introduced by Bruce Sterling \cite{sterling2005press} and is defined as \textit{``the deliberate use of diegetic prototypes to suspend disbelief about change"} \cite{coulton2017design}.  Design Fiction serves as a valuable method in HCI by allowing researchers and designers to explore speculative futures and potential technological developments through narration and prototyping. This approach encourages critical thinking about the social, ethical, and cultural implications of emerging technologies, fostering innovation while grounding design in real-world contexts \cite{grand2010design,coulton2017design,lindley2016pushing}. By envisioning and engaging with possible futures, Design Fiction helps to anticipate challenges, inspire novel solutions, and provoke discussions that can shape more thoughtful and human-centered technological advancements. This approach thus helps to surface potential issues and challenges that might not be evident through traditional research methods, fostering a more comprehensive understanding of the interplay between technology and society \cite{lindley2016pushing}. 


Design fiction has been used in different contexts \cite{coulton2017design,blythe2018research,blythe2014research}. \cite{muller2020understanding} suggest, \textit{``Design fictions are increasingly important and prevalent within HCI, though they are created through diverse practices and approaches and take many diverse forms"}. Regarding outcomes of Design Fiction, \cite{blythe2014research}suggests that \textit{``Design fictions can take the form of narratives, short stories, films but also objects and semi-working prototypes"} . Design Fiction operates not merely as a forecasting tool but as a medium for generating reflective and discursive spaces where the implications of emerging technologies can be interrogated. \cite{blythe2014research} asserts, \textit{``Fiction has also been used critically to re-consider existing or emerging ubiquitous computing technology."}. For instance, Design Fiction can challenge assumptions and inspire innovation within HCI by presenting alternative design possibilities that provoke critical discourse \cite{auger2013speculative}. Further, design fiction can serve as a lens for understanding user experience by creating fictional prototypes that reveal user interactions in imagined contexts, thereby exposing potential usability issues and ethical considerations \cite{blythe2014research}. Design Fiction has also been used in more `practical' settings like critiquing smart city technologies. By crafting fictional scenarios that depict the integration and impact of these technologies in urban environments, designers can critically assess the social and ethical ramifications before actual implementation \cite{coulton2017design}. \cite{lindley2016pushing} pushing the limits of design fiction state,  \textit{``Imaginary abstracts summarize the findings of papers that “have not been written” and report on “prototypes that do not exist”. The general premise “might be a shocking argument” but “perhaps fictional user studies might be a means of reflecting on what might be learned through prototype development”''}. 

This work is both inspired by and built upon the success and acceptance of design fiction as a methodological approach and a research strand \cite{grand2010design} within HCI.

\subsection{Thought Experiments in Science and Design Fiction}

Thought experiments have long been a cornerstone of scientific inquiry, providing a method for exploring hypothetical scenarios and testing the implications of theoretical concepts without the need for physical experimentation \cite{sep-thought-experiment}. Also known as \textit{gedanken experiment} in the natural sciences, they allow scientists to take existing information that based on known phenomena and mentally manipulate it into new configurations that can advance the frontiers of knowledge \cite{mccomb2012thought}.  Science fiction has historically harnessed thought experiments to explore complex scientific and philosophical questions through imaginative narratives. Pioneering authors like Isaac Asimov \cite{meillassoux2015science} and Philip K. Dick \cite{huntington1988philip} employed thought experiments to probe the consequences of advanced technologies and societal changes in their works. For instance, Asimov's "Robot" series extensively used the Three Laws of Robotics \cite{mccauley2007ai} as a thought experiment to investigate ethical dilemmas and the relationship between humans and artificial intelligence. Similarly, Dick’s exploration of altered realities and identity in "Do Androids Dream of Electric Sheep?" \cite{dick2014androids} reflects speculative inquiry into the nature of consciousness and human-machine interaction. These speculative scenarios serve not merely as entertainment but as a means to reflect on the implications of technological advancements and human values. 

In the realm of HCI, the utility of thought experiments extends beyond traditional scientific boundaries into the domain of design fiction. Design fiction, which blends speculative design with narrative storytelling, leverages these hypothetical constructs to envision future technologies and their impacts on human interaction. Blythe and Encinas \cite{blythe2018research}, characterize thought experiments as a fundamental aspect of design fiction, serving as a crucial method for exploring and conceptualizing possible futures within the design discipline. Thought experiments are described as imaginative exercises that allow designers and researchers to explore scenarios that may not be feasible or practical to test in reality \cite{blythe2018research}. This approach is aligned with practices in other fields such as physics, mathematics, ethics, and philosophy, where thought experiments are used to probe the implications of theories and hypotheses without empirical testing.  In our work, we adopt and build upon this characterization of thought experiments.

\subsection{Design Fiction for Visualization}

Design fiction, as a methodological approach, can offer a unique lens through which the field of visualization can be both critically examined \cite{gray2016ways,naerland2020political} and have its scope expanded \cite{correll2018ross,owen2024design}. At its core, design fiction involves the creation of speculative scenarios and artifacts that challenge current assumptions,  propose alternative methods and thus provide a criticism for visualization. Kosara \cite{kosara2007visualization} argues that \textit{``Expertise in a professional critique means anticipating, as much as possible, all potential concerns for the design, from purely practical issues to matters of content and even broader social implications."}. On the importance of visualization criticism Kosara further suggests that,  \textit{``This (the act of visualization criticism) is no different from designing successful visualization techniques."}. \cite{drucker2017information} asserting visualization as enunciation suggests, \textit{``A grid, graph, schematic tree, chart, or diagram is a cultural construct, no matter how much its visual qualities might intuitively suggest analogies with natural phenomena or processes."}. And as every cultural construct, visualization is subject to criticism \cite{poovey1990cultural}.

\section{Anatomy of the Thought Experiments}

We define the thought experiments (\textit{TE}) as consisting of a situation, story and a scenario.
\begin{equation*}
    ThoughtExperiment (\bm{TE}): \{SITUATION, STORY, SCENARIO\}
\end{equation*}

The structure of the thought experiments has been deliberately crafted to incorporate elements of design fiction in visualization. We define each of these components as; 
\begin{enumerate}
    \item \textit{\uppercase{Situation}}: A situation is basically derived from the tools of the thought experiments and from reflections on visualization practice. 
    \item \textit{\uppercase{Story}}: The story is basically a short fiction \cite{blythe2017research}. In this work, these short fictions are generated by AI.
    \item \textit{\uppercase{Scenario}}: The scenario is based on the situation and the generated story while being grounded in visualization research.  
\end{enumerate}

The situation is used to generate the story, and the situation and the story is used to develop the scenario. The relationship between each of the components is shown in the diagram below.

$$
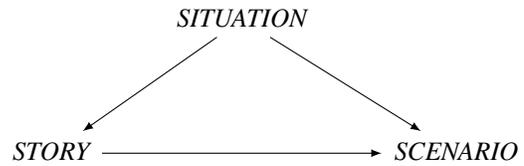
\begin{tikzcd}
              & {\textit{\uppercase{Situation}}} \arrow[ldd] \arrow[rdd] &    \\
              &                            &    \\
{\textit{\uppercase{Story}}} \arrow[rr] &                            & {\textit{\uppercase{Scenario}}}
\end{tikzcd}
$$


\subsection{Situations}
Situations are developed using different thought experiment tools and from reflecting upon visualization practice. 
\subsubsection{Thought Experiment Tools} \label{toolkit}

We chose to use three tools of thought experiments, counterfactuals \cite{lebow2007counterfactual}, retrodiction \cite{jones2007mind} and backcasting\cite{coric2021importance}. 

\begin{enumerate}
    \item \textbf{Counterfactuals} in thought experiments refer to hypothetical scenarios or conditions that deviate from actual empirical observations or historical events \cite{lewis2013counterfactuals,lebow2007counterfactual}. These speculative constructs serve as crucial components of thought experiments, enabling researchers to explore the consequences of altering particular variables or circumstances within a given theoretical framework. By positing counterfactuals, researchers can explain causal relationships, evaluate the robustness of theories, and assess the potential outcomes of alternative scenarios. Counterfactual reasoning entails the mental simulation of events or conditions that did not occur in reality but are conceptually plausible within the confines of a particular hypothesis or theoretical framework. Through the manipulation of counterfactuals, researchers can uncover hidden assumptions, challenge conventional wisdom, and generate new insights into the underlying mechanisms governing natural phenomena or social dynamics \cite{tetlock1996counterfactual}.

    \item \textbf{Retrodiction} within the context of thought experiments involves the retrospective application of theoretical principles or models to past events or empirical data \cite{jones2007mind}. Unlike prediction, which entails forecasting future outcomes based on existing knowledge or theoretical frameworks, retrodiction involves reconstructing past events or conditions by employing established theories or conceptual frameworks. Through retrodictive analysis, researchers aim to assess the explanatory power of their theories by demonstrating their ability to account for observed phenomena or historical data retrospectively. By scrutinizing the consistency between theoretical predictions and known historical outcomes, retrodiction serves as a critical evaluative tool for assessing the validity and explanatory adequacy of scientific theories. Moreover, retrodiction allows researchers to uncover potential limitations or inconsistencies within theoretical frameworks and refine their models to better align with empirical observations \cite{philip1996counterfactual}. 

    \item \textbf{Backcasting}, within the realm of thought experiments, denotes a methodological approach wherein researchers work backward from a desired future outcome to identify the necessary conditions, actions, or interventions required to achieve it \cite{holmberg2000backcasting}. Unlike forecasting, which projects future states based on current trends or historical data, backcasting entails envisioning a desirable future state or goal and then delineating the steps or strategies needed to realize it. Rooted in the principles of normative reasoning and goal-oriented planning, backcasting serves as a forward-thinking tool for designing policies, strategies, or interventions aimed at addressing complex societal challenges or achieving specific objectives \cite{hojer2000determinism,robinson1982energy}. By envisioning a preferred future scenario and systematically tracing the pathways to reach it, backcasting empowers researchers and policymakers to explore alternative trajectories, anticipate potential obstacles, and devise innovative solutions.

\end{enumerate}
\subsubsection{Reflections on Visualization Practice}

Reflecting on visualization practice involve examining various visualization case studies through discussions with industry-based  visualization practitioners. The case studies involved different challenges, insights, or innovative approaches to visualization that have emerged from the author's interactions. Through these experiences and interactions, the authors gained valuable insights into the diverse approaches to visualization, the challenges that practitioners encounter and the different complexities that underlie the visualization process. By examining some key aspects of these case studies, we were able to identify some themes that we could integrate with the different tools of the thought experiments. 

\subsection{Stories}

Stories are included in the thought experiments to make situations more personal and contextual, thereby enhancing their user relatability and depth. In this work, we use AI to generate the stories. By using AI, we aim to amalgamate various contexts to identify and explore scenarios where visualization can offer significant insights and advantages.


\subsubsection{AI as a storyteller}
AI systems such as Large Language Models (LLMs) exhibit substantial storytelling capabilities \cite{zhao2023more}, attributable to their extensive training on diverse and voluminous textual datasets, which include various narrative forms such as novels, short stories, and other literary compositions.  By leveraging their ability to predict word sequences, LLMs can construct complex plots, develop characters with depth, and create immersive settings, all while maintaining thematic consistency and adapting to specific prompts  that guide the narrative’s direction, tone, or emotional undertone. 

\subsubsection{Prompting}
We utilized the GPT-4o LLM model with zero shot prompting. Zero-shot prompting refers to the technique of providing LLMs with a question without any prior examples or specific instructions, allowing the LLM to generate a response based solely on its pre-existing knowledge.  

We prompted GPT-4o to generate a story based on a the theme. Specifically, the prompt was to generate a story in about 200 words about the given situation.
\subsection{Scenarios}
Scenarios are basically short essays that we develop by grounding the situation and the story in visualization research. We find concepts in visualization which closely capture the ideas in the situation and the story. We then contextualize situations, try to relate the stories and quote relevant work in order to ground them in visualization research.

\section{Thought Experiments for Visualization}

The following thought experiments (TE1-TE10) have been derived using the methods outlined above. TE1-TE4 are based on counterfactuals, TE5-TE7 are based on retrodiction and TE8-TE10 are based on backcasting. 

\subsection{\textbf{TE 1}} 

\begin{enumerate}
    \item \textit{\uppercase{Situation}}: If more data would have been there, the visualization and inference that the users derive from the visualization would likely be significantly different
    \item \textit{\uppercase{Story}}:\texttt{In a bustling urban planning office, a team was tasked with designing a new public park. They had access to a limited dataset that included basic demographics and current green spaces in the area. With this data, they created a series of visualizations: bar graphs, pie charts, and heat maps that illustrated the distribution of existing green spaces and projected usage patterns. The initial visualizations suggested a surplus of park space in affluent neighborhoods and a deficit in underdeveloped areas. Based on these findings, the team proposed placing the new park in a wealthy district, assuming that the current data accurately represented community needs. However, a deeper analysis revealed that the dataset was incomplete. Key information, such as socioeconomic trends, environmental impact assessments, and detailed neighborhood surveys, was missing. Had this additional data been available, the visualizations would have highlighted a more nuanced picture: underrepresented community needs, environmental concerns, and accessibility issues that were not apparent with the initial dataset. In the end, the team recognized that while their initial visualization was informative, the decision-making process could have been significantly improved with a more comprehensive dataset, leading to a park that better served the entire city.}
    \item \textit{\uppercase{Scenario}}:  In many instances, visualizations are constructed using incomplete datasets, which may arise from various sources of data incompleteness. This may include missing data, both known and unknown, data that cannot be captured, or data that has not been identified or considered relevant. The quantity of available data plays a critical role in determining the quality and reliability of the visualizations and the inferences drawn from them. Insufficient or incomplete datasets can distort the representation of phenomena, often resulting in biased, skewed, or overly simplistic visualizations. Visualization research has taken into account missing data\cite{song2018s,swayne1998missing,song2021understanding,fernstad2019identify}. Missing data has been treated as black holes in visualization \cite{song2018s}. Missing data  generally referees to formalize and those that can be filled with measures of central tendency, or some domain heuristics \cite{song2018s}. Furthermore, small datasets may fail to capture complex relationships between variables, increasing the risk of drawing erroneous conclusions. Visualization research has also been very active in dealing with uncertainty in data \cite{brodlie2012review,hullman2019authors}. \cite{hullman2019authors} in interviews found that \textit{``many authors also said they would like to communicate uncertainty more frequently, and expressed intuitions that doing so would shift people’s general tolerance for and ability to reason with uncertainty."}. But this is when the authors know about uncertainty in data (the known unknown). But what about the unknown unknown \cite{wagenmakers2023facing}? How to capture that in a visualization? Or is it out of scope of visualization? The team that was visualizing data they were unaware that there was another dataset that they did not consider. Some times the other dataset can make a difference. Other times the dataset only makes the difference: making the the entire visualization pointless. \cite{wagenmakers2023facing} suggest that \textit{``most uncertainty may lie not in what is usually reported (e.g., p value, confidence interval, or Bayes factor) but in what is left unreported (e.g., how the experiment was designed, whether the conclusion is robust under plausible alternative analysis protocols, and how credible the authors believe their hypothesis to be). "}. The \textit{unreported} can contain plenty of other factors, and that depends on the problem under consideration. 
   
\end{enumerate}

\subsection{\textbf{TE 2}}

\begin{enumerate}
    \item \textit{\uppercase{Situation}}: If the metadata had been included in the visualization, the resulting inference could have been more specific and thus more correct.
    \item \textit{\uppercase{Story}}:\texttt{In the heart of a research lab, a team of data scientists was analyzing climate change trends using a sophisticated visualization tool. Their primary dataset included temperature readings from various global stations over the past decade. The visualizations they created—line graphs and scatter plots—displayed significant warming trends across multiple regions. However, the team chose not to include metadata such as measurement methods, station calibration details, and temporal resolution in their visualizations. The absence of this crucial information led to a generalized interpretation of the data. The visualizations suggested uniform warming patterns across all regions, which seemed consistent with broad climate change theories. Months later, a peer review highlighted the omission of metadata as a critical oversight. The review revealed that certain stations had varying calibration standards and that some regions had more frequent data gaps than others. Incorporating this metadata would have allowed for a more granular analysis, revealing that regional warming trends were not as uniform as initially thought. Some areas experienced faster warming due to local factors, while others had slower rates due to incomplete data coverage. With the metadata included, the team could have produced a more accurate and specific inference, leading to refined climate models and targeted mitigation strategies.}
    \item \textit{\uppercase{Scenario}}:  The inclusion of metadata in visualizations \cite{burns2022invisible} significantly influences the interpretation and subsequent inferences drawn from the data. Metadata, which provides essential context such as the origin, methodology, and limitations of the data, enables a more nuanced and better understanding of the visual representation \cite{baca2016introduction}. \cite{burns2022invisible} concluded from their study that \textit{``We found that visualizations with metadata were perceived as more thorough, but similarly accurate, clear, complete, and relevant in comparison to visualizations without metadata."}.  Without this contextual information, viewers might rely solely on (external and internal) visual cues \cite{kong2017internal}, potentially leading to oversimplified or incorrect conclusions. Metadata could also mean including the data bias, limitations of the data or some information within the domain. This would actually help giving the user context behind the data. For instance, a graph displaying a correlation between two variables without metadata may lead to the assumption of causation. However, if the metadata indicates that the data were collected from a non-random, convenient  sample or under specific conditions, the viewer might recognize that the correlation does not imply causation, thereby preventing erroneous inferences. Furthermore, metadata can elucidate the scope, context and applicability of the data, influencing how findings are generalized. Climate change is a contentious issue \cite{AP_2023}, with varying opinions among individuals.Consequently, if data collection is biased or involves selective reporting, the interpretation of visualizations must be adjusted accordingly, underscoring the importance of metadata.  Further, metadata can highlight any pre-processing steps, such as data cleaning, transformation, or missing data analysis which could affect the data’s integrity. In the absence of this critical information, viewers may either overestimate the robustness and accuracy of the visualization or misinterpret anomalies. This concern is particularly relevant when data originates from political parties or sources with vested interests.

\end{enumerate}

\subsection{\textbf{TE 3}}

\begin{enumerate}
    \item \textit{\uppercase{Situation}}: Had the user been properly trained in reading and interpreting data visualizations, they could potentially yield different insights.
    \item \textit{\uppercase{Story}}:\texttt{In a high-tech financial firm, analysts regularly used
complex data visualizations to track market trends
and forecast economic conditions. One day, a new team
member, Alex, was assigned to review the quarterly
performance reports. The visualizations included intricate
line charts, heat maps, and bubble plots depicting stock
performance, trading volumes, and economic indicators.
Alex, though highly skilled in finance, had not received
formal training in data visualization. As a result,
he struggled to interpret the nuanced patterns in the
visualizations. His analysis concluded that the market
was experiencing only minor fluctuations and recommended
maintaining the current investment strategy. However, a
more seasoned analyst, Emma, who had undergone extensive
training in data visualization, reviewed the same visualizations.
She noticed subtle trends and correlations that Alex had
missed, such as cyclical patterns in trading volumes
and emerging anomalies in economic indicators. Emma’s
interpretation revealed that the market was on the
verge of a significant shift, prompting a strategic
realignment. Had Alex received proper training in reading
and interpreting data visualizations, he might have
detected these critical insights, potentially altering
the firm’s investment approach and avoiding a financial
setback.}
    \item \textit{\uppercase{Scenario}}:  Visualization literacy is a well explored and researched theme in visualization research \cite{borner2019data,boy2014principled}. Accurate interpretation of data visualization is a learned skill — involving discernment of complicated chart-types, understanding the contextual relationship between data-sets, and identifying potential bias in the representation \cite{suciu2021data}. A user well versed in these aspects can extract different insights than an untrained individual. Training and education in data visualization bolster users' ability to unravel intricate stories data might depict and identify underlying patterns or anomalies, leading to different and potentially more accurate insights \cite{Iliinsky2012properties}. The more adept users are at interpreting visualisations, the more successfully they can leverage this tool for in-depth analysis \cite{borner2019data,boy2014principled}, leading to varied, comprehensive and potentially transformative insights. Alex needs some training in interpreting (ideally financial)  charts. Emma, a more seasoned analysis, probably had better visualization literacy.

\end{enumerate}


\subsection{\textbf{TE 4}}

\begin{enumerate}
    \item \textit{\uppercase{Situation}}: If the user has prior knowledge or expectations from the data, this might influence the interpretation of our visualization, leading to a potentially incorrect conclusion. 
    \item \textit{\uppercase{Story}}:\texttt{In a pivotal election cycle, a political analyst named
Sam was tasked with reviewing a series of data visualizations
depicting voter turnout and election results across
various regions. The visualizations included bar charts
showing turnout rates, pie charts detailing party preferences,
and heat maps indicating voting patterns. Sam, a committed
supporter of a particular political party, approached
the data with a clear expectation: that their preferred
party would show stronger performance in key battleground
areas. As he reviewed the visualizations, Sam’s prior
knowledge and expectations influenced his interpretation.
For instance, when he saw a heat map with several areas
shaded in a lighter color, he interpreted this as a sign
of growing support for his party, even though these
areas had lower turnout rates overall. In contrast,
an unbiased analyst reviewing the same visualizations
might have noted that the lighter shades represented
regions with low voter engagement and that the actual
performance data did not support the anticipated surge
in support for Sam’s party. The unbiased analyst would
have identified that high turnout areas, which were
crucial for determining true support levels, did not
show a significant shift. Sam’s pre-existing expectations
led to a skewed interpretation of the data, potentially
resulting in an incorrect conclusion about the party’s
performance and misinforming campaign strategies. Had
he approached the visualizations with a neutral perspective,
the analysis would have provided a more accurate understanding
of voter dynamics.}
    \item \textit{\uppercase{Scenario}}:  The interaction between user's prior knowledge and the visualization of data has a significant bearing on visualization inference, and thus the user can be prone to misinterpreting the visualization \cite{borner2019data, boy2014principled}. Understanding the influence of prior knowledge or expectations of the user is key to effective communication of the data \cite{kirk2016data}. A user's prior knowledge, cognitive biases, or expectations can significantly influence their perception and interpretation of data visualizations. This cognitive state often affects how users understand, interpret, and draw conclusions from visualized data, potentially leading to deviations from the actual or intended implications of the data. For instance, political contexts are inherently divisive, and this divisiveness can be reflected in the misrepresentation of data through visualizations. \cite{cairo2019charts} suggests political contexts often exhibit clear distortions in graphical representations. But Cairo primarily discusses this from the designer's perspective. Users can also willfully ignore certain data or fail to recognize errors in visualization design. While an unbiased user might identify errors such as selective data presentation, manipulation of visual elements, or omission of critical information, a biased user may overlook these errors when a favorable or expected pattern is present in the visualization. Sam's commitment to a particular political party made him interpret lighter colors incorrectly.

\end{enumerate}

\subsection{\textbf{TE 5}}
\begin{enumerate}
    \item \textit{\uppercase{Situation}}: The designer developed \& designed the visualization incorrectly
    \item \textit{\uppercase{Story}}:\texttt{In a health analytics firm, a data visualization designer,
Maria, was assigned to create a dashboard for tracking
patient outcomes from various treatments. Her goal
was to design an intuitive and informative set of
visualizations to help medical professionals make data-driven
decisions. Maria developed several visualizations, including
a series of line graphs and pie charts. However, due
to a series of errors in her design process, the
visualizations were misleading. For example, the line
graphs used inconsistent scales, which distorted the
trends in patient recovery rates. Additionally, the
pie charts were poorly labeled, with some sections
overlapping, making it difficult to discern the exact
proportions of different treatment outcomes. When the
dashboard was rolled out, medical professionals struggled
to interpret the data accurately. The inconsistent
scales led to incorrect conclusions about the effectiveness
of different treatments, while the confusing pie charts
obscured critical information about treatment success
rates. After feedback from users, it became clear that
the issues were not with the data itself but with
Maria’s design choices. The visualizations did not
effectively communicate the underlying data, highlighting
the importance of accurate and thoughtful design in
data presentation. Maria revised the visualizations to
correct the scales and improve labeling, resulting in
a more accurate and user-friendly dashboard.}
    \item \textit{\uppercase{Scenario}}:  The visualization designer developed an incorrect visualization due to a failure to adhere to fundamental principles of effective data visualization, leading to a misleading representation of the data \cite{lo2023change,bigelow2014reflections}. Maybe the choice of chart type was inappropriate for the nature of the data being presented. For instance, using a pie chart to display data that requires a comparison over time, which is better suited to a line graph or bar chart, can obscure the intended message and lead to incorrect inferences. The designer may have neglected to maintain consistent scaling of the axes, which can distort the viewer's perception of the data's magnitude and trends. This inconsistency can create the illusion of significant changes where none exist or mask important variations \cite{long2024cut}, ultimately compromising the integrity of the visualization. The visualization may have suffered from poor use of color and visual hierarchy, which are crucial for guiding the viewer's attention to the most important aspects of the data. The overuse of bright colors \cite{silva2011using,szafir2017modeling} and the lack of contrast can overwhelm the viewer and obscure key data points. Furthermore, the absence of clear labeling and legends could leave the audience struggling to understand the context and meaning of the visualization. This lack of clarity can lead to misinterpretation, as viewers might rely on their assumptions or cognitive biases \cite{dimara2018task} to fill in the gaps. The designer may have also failed to provide sufficient context or background information, which is essential for accurate data interpretation. Without understanding the data's source, methodology, and limitations, viewers are left to draw conclusions based on incomplete information. These design flaws collectively contributed to the ineffective and potentially misleading nature of the visualization, highlighting the importance of adhering to established best practices in data visualization to ensure accurate and meaningful data representation. It is also possible that there was a deliberate attempt from the designer to mislead the audience. The designer may have engaged in employing dark patterns \cite{gray2018dark,lukoff2021can}. This has been seen in many context, though difficult to evidence and therefore characterize. Maria, assuming having good intentions, needed to learn principles of effective data visualization. 
\end{enumerate}

\subsection{\textbf{TE 6}}
\begin{enumerate}
    \item \textit{\uppercase{Situation}}: The device distorted the visualization
    \item \textit{\uppercase{Story}}:\texttt{In a high-profile urban planning project, a team of architects and city planners relied on a sophisticated data visualization device to analyze traffic flow and pedestrian movement in the city center. The device projected a 3D model of the cityscape onto a large interactive screen, allowing the team to assess traffic patterns and make informed decisions about infrastructure changes. During a crucial meeting, the device malfunctioned. The projection was distorted, with significant misalignment in the 3D model and inaccurate scaling of traffic flows. For instance, major traffic bottlenecks were displayed in incorrect locations, and pedestrian pathways appeared distorted, leading to a skewed understanding of actual movement patterns. Despite the team’s expertise, the device’s malfunction led to misinterpretations of the data. Recommendations were made based on the flawed visualizations, such as rerouting major traffic lanes and redesigning pedestrian crossings, which did not address the real issues. It was only after the device was repaired and the correct visualizations were reviewed that the team realized the extent of the distortion. The insights drawn from the faulty device had led to potentially costly design errors. This incident underscored the critical importance of ensuring the accuracy and reliability of visualization devices in data-driven decision-making processes.}
    \item \textit{\uppercase{Scenario}}:  The choice of device used to display a visualization can significantly impact its effectiveness and potentially distort the conveyed information. Variations in screen size, resolution, and color calibration across devices can alter the visual appearance and interpretability of data. For instance, a visualization designed for a high-resolution monitor may lose detail or become unreadable when viewed on a smaller, lower-resolution screen, such as a smartphone, if it not designed and developed correctly. \cite{horak2021responsive} assert that \textit{``the visualization community has yet to establish responsive design principles to a similar (web design community) extent."}. They further add, \textit{``these existing (visualization) techniques ( for smartphone, tablet, and watch interfaces) tend to focus on single contexts rather than on varied and dynamic usage contexts"}. Furthermore, differences in touch interaction versus mouse input can affect the user's ability to explore the data, leading to possible misinterpretations. The physical environment in which a device is used, such as lighting conditions or the level of ambient light, may also influence how colors and contrasts are perceived, further complicating the accurate interpretation of visual information. Consequently, ensuring that visualizations are adaptable and responsive to various device characteristics is crucial for maintaining the integrity of the data and avoiding unintentional distortion. In the urban planning project, the device malfunctioned by projecting the image; when the correct visualization required an accurate projection. For the urban planning project, the distortion wasn't within the scope of visualization research. 
\end{enumerate}

\subsection{\textbf{TE 7}} 

\begin{enumerate}
    \item \textit{\uppercase{Situation}}: The data was incorrect to begin with; visualization couldn't have done anything to prevent misread or misinterpretation
    \item \textit{\uppercase{Story}}:\texttt{In a public health organization, a team was tasked
with analyzing and presenting data on vaccination rates
across different regions. They utilized an advanced
visualization tool to create a series of interactive
charts and maps, aiming to provide clear insights into
regional vaccination coverage and identify areas in
need of intervention. However, the data feeding into
the visualization was fundamentally flawed. A data
entry error had resulted in incorrect vaccination rates
being recorded for several key regions. For instance,
data for Region X was mistakenly recorded as having a
90
to 50
visualizations were completed.The visualizations displayed
these erroneous figures, with maps highlighting Region
X as having exceptionally high vaccination coverage
and charts indicating overall strong performance in the
region. Despite the clear and user-friendly presentation
of the data, the visualizations could not compensate
for the underlying inaccuracies. Stakeholders were led
to believe that intervention was unnecessary in Region
X, while in reality, it was an area in critical need of
vaccination efforts. It was only after a thorough audit
revealed the data discrepancies that the limitations
of the visualization became apparent. The incident
demonstrated that no matter how effective the visualization
tool, the integrity of the data remains paramount. The
visualization’s ability to convey accurate information
is directly dependent on the accuracy of the data it
represents.}
    \item \textit{\uppercase{Scenario}}: The importance of reliable, accurate, trustworthy data cannot be overstated, as it forms the basis for all inferences, decisions, actions across all disciplines. Reliable, accurate, trustworthy data is also thus required for reliable, accurate and trustworthy visualizations. In spite of the most sophisticated visualization techniques and methods in our toolkit, these tools are inevitably rendered impotent when faced with distortions in the foundational raw data. Visualization functions as a lens through which we view and interpret data, and just like any lens, it can only enhance what exists - it cannot rectify inherent distortions in whatever it is observing, measuring or quantifying \cite{tufte1990data}. Fabricated data leads to fabricated visualization. This underscores the importance of rigorous data quality assurance procedures (particualrly in high stakes domains) and the conduct of appropriate and adequate data pre-processing prior to visualization. However, visualization techniques can be useful for detecting some kinds of errors \cite{wegman1990hyperdimensional}. For instance, exploratory visualizations might reveal anomalies, outliers, or patterns that suggest issues with the underlying data, serving as a catalyst for further investigation \cite{keim2002information}. If inaccurate or mistakenly recorded data is processed, even the most accurate visualization tools are rendered futile and can lead to widespread misinformation and thus misguided decisions. As such, it is of utmost importance to implement stringent quality control measures for data collection and preparation stages to ensure robust and reliable visualizations that users can trust and make decisions based on it \cite{kimball2008data}. Consequently, educating users about the role and significance of data quality \cite{wolff2016creating} in visualization, not only to interpret but also to recognize potential signs of data inaccuracies or quality issues in visualizations is a key aspect in increasing awareness and promoting a critical and informed attitude. Correct vaccination rates are required for a correct visualization.
\end{enumerate}

\subsection{\textbf{TE 8}} 

\begin{enumerate}
    \item \textit{\uppercase{Situation}}: The user is going to pick up a trend which does not exist so there is a requirement of diligence with the design choices used in the visualization.
    \item \textit{\uppercase{Story}}:\texttt{In a market research firm, analysts were tasked with
evaluating consumer behavior trends over the past year
using a new visualization tool. The tool was designed
to display various metrics, such as sales figures and
customer demographics, through line graphs and bar
charts. The lead analyst, Jamie, had a preconceived
hypothesis that a certain product category was experiencing
a significant upward trend in consumer interest. The
visualization tool, however, included a design flaw:
the line graphs used inconsistent intervals for data
points, which unintentionally exaggerated fluctuations
and created the appearance of a trend where none existed.
For example, a slight increase in sales was visually
amplified due to irregular spacing between data points.
Jamie, focused on validating the hypothesis, noticed
these exaggerated fluctuations and concluded that the
product category was indeed seeing a substantial rise
in interest. This conclusion, based on the distorted
visualization, led to a recommendation for increasing
production and marketing efforts for that category.
Only after a more detailed analysis, which involved
correcting the design flaw and using consistent data
intervals, did the team discover that the supposed trend
was a result of visualization artifacts rather than
an actual market shift. The incident highlighted the
critical need for diligence in design choices to ensure
that visualizations accurately represent the data and
prevent the misinterpretation of non-existent trends.}
    \item \textit{\uppercase{Scenario}}:  Careful consideration of design choices \cite{tang2004design} is important to prevent incorrect visualization inference.This is crucial in cases where users perceive patterns or trends within data that are do not exist at all or are not statistically significant. This bias\cite{dimara2018task} underscores the necessity for meticulous design strategies aimed at facilitating accurate interpretations of visualized data. While the human propensity to discern patterns is a fundamental aspect of cognition, it also renders users susceptible to identifying false correlations or trends \cite{lo2022misinformed}, particularly in complex datasets. Moreover, the absence of rigorous design considerations may inadvertently facilitate the emergence and perpetuation of spurious trends within the data. In instances where users encounter data visualizations devoid of clear contextualization or comprehensive explanatory frameworks, there exists a heightened likelihood of misinterpretation or the identification of false trends \cite{lisnic2023misleading}. Such misinterpretations can lead to misguided conclusions, potentially influencing decision-making processes or public perceptions \cite{doan2021misrepresenting}.  For examples, users have a propensity to assume causality from visualized data. As Xiong et al. \cite{xiong2019illusion} suggest \textit{``Concluding causality from the visualized data alone is misguided. We can only establish a correlation - the tendency of two variables changing together - between temperature and crime rate because it is possible that other factors not shown on the graph caused the difference in the number of violent crimes."}. Thus sufficient context is required to suggest the user that there may be other important variables not captured in the visualization.  Jamie, having a preconceived hypothesis, picked up a non existent trend because of poor design choices used in the visualization. 
\end{enumerate}

\subsection{\textbf{TE 9}}
\begin{enumerate}
    \item \textit{\uppercase{Situation}}: The user is going to make very influential inference from the visualization, so apart from being careful with design choices, it is important provide appropriate guidelines to the user to avoid major biases.
    \item \textit{\uppercase{Story}}:\texttt{In a financial consulting firm, a team was tasked with evaluating investment opportunities using a set of interactive data visualizations. These visualizations included complex scatter plots and trend lines designed to illustrate potential returns on various investment portfolios. The results would influence key decisions, including large-scale investments and strategic shifts for high-profile clients. Emma, a senior consultant, was preparing a report based on these visualizations to present to the firm’s executives. The visualizations were well-designed, incorporating detailed data points and sophisticated graphical elements. However, due to the high stakes involved, it was critical to ensure that the interpretations drawn from these visualizations were accurate and unbiased. The design choices included advanced filters and color coding, which could lead to misinterpretation if not used correctly. For instance, a color gradient intended to show risk levels could be misread if the user was not familiar with its scale. Additionally, the scatter plots displayed dense clusters of data points that required careful analysis to avoid overestimating trends. Recognizing the potential for influential inferences, the firm provided Emma with detailed guidelines on how to interpret the visualizations, including explanations of color schemes, scales, and data clustering. These guidelines were essential for ensuring that Emma’s analysis was grounded in the actual data, minimizing the risk of bias or misinterpretation. Despite these precautions, Emma's presentation of the visualization led to a significant decision to allocate a substantial portion of the firm’s investment portfolio to a specific sector. The guidelines helped Emma avoid major biases and ensured that the influential inference drawn from the visualizations was based on a careful and accurate interpretation of the data.}
    \item \textit{\uppercase{Scenario}}:  Visualization is crucial for user inference and decision-making in domains like medicine \cite{gillmann2021ten,johnsonchris2022review} or aerospace \cite{barladian2021safety}. These visualizations possess a profound capacity to shape users' perceptions, interpretations, and subsequent actions \cite{lauer2020people}. Consequently, it is important to exercise caution and prudence in the creation and dissemination of visualizations to mitigate the potential for biases or errors to influence users' inferences. For example visualization in a prominent newspaper for a large audience right before an election has the potential of influencing a voting decision \cite{mclaughlin2019react}. To address the threat of biased inference stemming from visualizations, it is imperative to establish and adhere to comprehensive guidelines governing the creation, presentation, and interpretation of visual data. Such guidelines should encompass principles of transparency \cite{weissgerber2019reveal}, accuracy \cite{munzner2009nested}, and user empowerment, thereby empowering users to critically evaluate visualizations and discern the underlying truths from potential distortions.  Moreover, educational initiatives aimed at enhancing users' visual literacy \cite{borner2019data,boy2014principled} and awareness of cognitive biases are indispensable in fostering a culture of discernment and skepticism in visual data representations. This is more critical in some high stakes domain such as medicine and aerospace. It is essential that these visualizations anticipate all potential outcomes and provide clear guidelines for their interpretation. This may include sample demonstrations that illustrate what different visual elements/instances within the visualization represent. For example, in flight deck data visualization, it is critical to precisely define the meaning of each visual instance to ensure clarity and prevent misinterpretation. In both fields, the accuracy and comprehensibility of visual information are paramount, as misinterpretation can lead to serious consequences. Medical professionals rely on visual data to make informed decisions about patient care, while pilots and pilots use flight deck visualizations to navigate and ensure flight safety. Thus, guidelines must be meticulously developed, ensuring that users understand how to interpret the data correctly. Furthermore, designers must account for the different contexts in which these visualizations will be used, as real-time decision-making in high-stakes environments demands both clarity and precision. Emma was provided with comprehensive guidance on interpreting the visualization, as would be expected in a high-stakes financial firm.
\end{enumerate}

\subsection{\textbf{TE 10}}
\begin{enumerate}
    \item \textit{\uppercase{Situation}}: The designer lacks technical and contextual understanding of the data and thus cannot design the visualization within a proper context. It is thus important to involve the data collection/subject expert right from the beginning of the visualization design process.
    \item \textit{\uppercase{Story}}:\texttt{In a regional transportation agency, a designer named
Lucas was assigned to create a series of visualizations
to communicate traffic congestion patterns and proposed
infrastructure improvements to the public. The visualizations
were intended to support a public consultation process
for a major road expansion project. Lucas, skilled in
graphic design but with limited knowledge of transportation
data and its nuances, set out to design line charts,
heat maps, and bar graphs based on the provided data
sets. However, lacking a deep understanding of traffic
flow dynamics and the specific context of the data, his
design choices were problematic. For instance, he used
overly simplified color schemes that failed to capture
the complexity of peak congestion times, and the heat
maps lacked clear demarcations for critical congestion
thresholds. When the visualizations were presented to
the public, they led to confusion and misinterpretation.
The general public struggled to understand the key
issues and potential benefits of the proposed road
improvements. For example, the heat maps did not adequately
convey the severity of congestion in certain areas,
leading to a misunderstanding of the necessity for
the planned changes. Recognizing the issue, the agency
involved a transportation data expert to review and
refine the visualizations. The expert provided crucial
contextual insights, such as peak traffic patterns and
the impact of proposed changes, which guided Lucas in
redesigning the visualizations. With this collaboration,
the revised visualizations were more accurate and informative,
effectively communicating the data within its proper
context and aiding the public in making informed decisions
about the road expansion project.
}
    \item \textit{\uppercase{Scenario}}: Designers lacking technical and contextual understanding of the data, can face significant challenges in creating effective visualizations. Data is imbued with context, nuances, and complexities that require careful interpretation. Without a deep understanding of the data's origins and context, the conditions under which it was collected, and its inherent limitations, designers may inadvertently create visualizations that misrepresent or oversimplify the information. Such misrepresentations can lead to erroneous conclusions, undermining the credibility of the visualization and potentially leading to poor decision-making. For instance, in the field of climate science, a visualization depicting temperature changes over time might be misleading if the designer does not account for variations in data collection methods or seasonal effects. If a visualization fails to accurately reflect these contextual factors, it might exaggerate or downplay trends, leading to incorrect interpretations about the severity of climate change. To mitigate these risks, it is crucial to involve subject matter experts in the visualization design process from the outset. These experts bring essential insights into the data’s context, including its nuances and potential pitfalls, ensuring that the visualization accurately reflects the underlying information. Tory and Moller \cite{tory2005evaluating}, in the context of visualization evaluation, support this approach and suggest that \textit{``One possibility is to have experts evaluate early prototypes (formative evaluation), and then have end users evaluate a refined version (summative evaluation)"}. This early evaluation will lead to the discovery of incorrect interpretation of the data in its context. Collaboration between designers and data experts facilitates a more informed design process, where visualizations are not only aesthetically compelling but also contextually relevant and technically sound. This approach enhances the overall effectiveness of the visualization, making it a more reliable tool for communication, analysis, and decision-making. Lucas needed some help, probably from someone who has had experience of working with transportation data. 
\end{enumerate}

\section{Discussion \& Visualization Critiques}

The thought experiments presented here lack the empirical data that predominates in visualization research \cite{lam2011empirical,weiskopf2020vis4vis}. Nevertheless, we contend that these thought experiments hold value for several reasons. First, they offer theoretical frameworks that can guide future empirical research. By posing hypothetical scenarios and exploring their implications, these experiments allow researchers to conceptualize complex phenomena that may not yet be easily testable. This can lead to the development of novel hypotheses or methodologies that inform subsequent empirical studies. Second, thought experiments stimulate critical thinking and innovation. By encouraging researchers to challenge established assumptions and consider alternative perspectives, they promote intellectual flexibility, which can be crucial for advancing the field. And, while empirical data is indispensable for validating theories, the role of thought experiments in shaping and refining those theories should not be underestimated. They provide a means to engage with abstract concepts in a way that complements empirical research, fostering a more comprehensive understanding of the subject matter.

This paper focuses on the subject of visualization, yet it intentionally omits the inclusion of any visualization or any dataset. We have chosen this approach because we believe that the presence of visual elements or datasets may inadvertently influence or skew the interpretation of our contribution, particularly in the context of design fiction. Furthermore, the absence of visuals aligns with our aim to engage in a conceptual and imaginative discourse, which is a core aspect of design fiction. The decision, thus, to exclude visual elements is an intentional  methodological choice that underscores the speculative nature of the work. 

Some of the scenarios have overlapping, have similar themes and kind of relate to each other. This is because scenarios are derived from situations which are further based on reflections on visualization practice. Many of these reflections on visualization practice revolve around common themes such as visualization in high-stakes, influential or divisive topics or domains or the role of data quality and integrity in visualization process. These recurring themes have influenced the development of the scenarios, resulting in certain scenarios overlapping with one another.


\subsection{Visualization Critique from the Thought Experiments}
From the thought experiments, we could distinctly derive three different generic visualization critique. They are generic because they do not apply to any specific visualization. We mostly adhered to the process and rules stated by Kosara \cite{kosara2007visualization}. These critiques have a neutral voice - are not directed towards the researcher but discuss the work. If the role of the designer is discussed, it takes the most impersonal tone without making any personal remarks. The critiques are based on facts - are they are derived from thought experiments that contain situations \& scenarios that is grounded in visualization practice \& research respectively. There is no self promotion in the criticism and each of the critiques contains a clear goal providing some alternate solution and or resolutions. 

\begin{enumerate}
    \item \textbf{Data blindspots}: There will be data that cannot be collected or is not available to the researcher, designer for visualization. So, these blind spots in data often fall outside the scope of what visualization research can address. Solutions to this is almost always context and domain dependent. For example, approaching data—and consequently, visualization—as a guideline rather than a definitive tool in decision-making. In visualization-driven decision-making, it is then essential to clearly communicate the limitations and caveats associated with any inferences drawn. 
    \item \textbf{Designer Intentions}: Data visualizations, when based on cherry-picked data, pose significant risks of bias, particularly when designers have the intent to mislead. Visualization simplifies complex datasets, making it an influential tool for shaping public perception. However, selective data representation, skewed scales, or omission of contextual information can distort reality, subtly guiding viewers toward a desired conclusion. This is especially concerning in contentious areas like climate change, immigration, and politics, where biased visualizations can reinforce polarized opinions or spread misinformation. The ethical responsibility of data visualization designers is critical, as they must prioritize transparency, accuracy, and balance in their work. Misleading visualizations not only compromise the integrity of the information being presented but also undermine public trust in data-driven discussions. Viewers, too, must develop visualization literacy skills to evaluate visualized data effectively. Ensuring ethical standards in both data selection and visualization design is vital to fostering informed, balanced discourse on divisive issues.
    \item \textbf{User Dilemmas}:This consideration follows from the previous one but shifts to the user's perspective. Users inevitably possess biases, often driven by their desire to believe certain conclusions. As a result, the scope of visualization research in mitigating these biases is inherently limited. This is a critical factor to acknowledge when evaluating visualizations through user studies \cite{greenberg2008usability}, particularly in divisive contexts where ethical constraints prevent the collection of comprehensive data from participants. Such limitations must be taken into account while formulating experiments or studies for evaluating visualizations \cite{elmqvist2012patterns}. In promoting visualization literacy, it is essential to educate users without introducing biases, thus ensuring they remain impartial in drawing inferences.
  
\end{enumerate}

\subsection{Limitations \& Future Work}
One fundamental limitation lies in the subjective nature of thought experiments, wherein mental simulations are dependent on individual cognitive faculties and predispositions. Thus the outcomes of thought experiments may vary widely among different individuals, designers, researchers and other stakeholders in visualization. This can make the establishment of consistent and universally applicable visualization principles \cite{zuk2006heuristics} difficult. However, thought experiments can be effective in critiquing, evaluating visualizations or providing domain-based or contextual principles. 

The GPT 4o generated stories, of the thought experiments, offered diverse, personalized contexts, enhancing thought experiments by making them more relatable and contextually relevant. However, these stories lacked finer, deeper details of the visualization. This may be a result of how GPT 4o is trained and implemented. A fine-tuned version of GPT with enhanced visualization capabilities could offer richer, more immersive stories, complete with detailed examples and a stronger focus on visualization. This is not a significant limitation of the thought experiments, as the scenarios provide the requisite the visualization grounding; but stories from a more tailored LLM can make them richer and more attuned towards visualization.



\section{Conclusion}

In this paper, we proposed thought experiments in design fiction for visualization. We defined thought experiments as consisting of situations, stories and scenarios. Situations were derived from tools of thought experiments and reflections on visualization practice; stories were generated by AI based on the situation and scenarios were basically the grounding of the situation and the story in visualization research. We presented ten such experiments and derived three generic visualization critiques from these thought experiments.


\newpage

\bibliographystyle{ACM-Reference-Format}
\bibliography{sample-base}

\end{document}